\documentclass[useAMS,aas_macros,usenatbib]{mn2e}
\usepackage{epsfig,subfigure}
\usepackage{latexsym,amssymb}

\newcommand{\boldsymbol}[1]{\mbox{\boldmath{$#1$}}}
\newcommand{\boldnabla}{\boldsymbol{\nabla}}
\newcommand{\boldv}{\boldsymbol{v}}
\newcommand{\boldx}{\boldsymbol{x}}
\newcommand{\boldg}{\boldsymbol{g}}
\newcommand{\boldB}{\boldsymbol{B}}
\newcommand{\boldQ}{\boldsymbol{Q}}
\newcommand{\boldb}{\boldsymbol{\hat{b}}}
\newcommand{\boldk}{\boldsymbol{k}}
\newcommand{\boldzeta}{\boldsymbol{\xi}}

\newcommand{\sigmabar}{\bar{\sigma}^2}

\newcommand{\chipara}{\chi_{\parallel}}
\newcommand{\chiperp}{\chi_{\perp}}
\newcommand{\chirad}{\chi_{\rm rad}}
\newcommand{\chiiso}{\chi_{\rm cond}}
\newcommand{\chiradiso}{\chi_{\rm iso}}
\newcommand{\Msun}{{\rm M}_{\odot}}

\begin{document}

\title[Collisional Buoyancy Instabilities]{Buoyancy Instabilities in Degenerate,
  Collisional, Magnetized Plasmas}
\author[Chang \& Quataert]{Philip Chang$^{1,2}$\thanks{E-mail:
pchang@cita.utoronto.ca (PC); eliot@astro.berkeley.edu (EQ)} \& Eliot Quataert$^1$\\
$^1$Department of Astronomy, 601 Campbell Hall, University of California, Berkeley, CA
  94720, USA\\
$^2$Canadian Institute for Theoretical Astrophysics, 60 St George St, Toronto, ON M5S 3H8, Canada
}

\maketitle 

\begin{abstract}
  In low-collisionality plasmas, anisotropic heat conduction due to a
  magnetic field leads to buoyancy instabilities for any nonzero
  temperature gradient.  We study analogous instabilities in
  degenerate {\it collisional} plasmas, i.e., when the electron
  collision frequency is large compared to the electron cyclotron
  frequency. Although heat conduction is nearly isotropic in this
  limit, the small residual anisotropy ensures that collisional
  degenerate plasmas are also convectively unstable independent of the
  sign of the temperature gradient.  We show that the range of
  wavelengths that are unstable is independent of the magnetic field
  strength, while the growth time increases with decreasing magnetic
  field strength.  We discuss the application of these collisional
  buoyancy instabilities to white dwarfs and neutron stars.  Magnetic
  tension and the low specific heat of a degenerate plasma
  significantly limit their effectiveness; the most promising venues
  for growth are in the liquid oceans of young, weakly magnetized
  neutron stars ($B \lesssim 10^9$ G) and in the cores of young, high
  magnetic field white dwarfs ($B \sim 10^9$ G).

\end{abstract}


\section{Introduction}\label{sec:introduction}

The presence of anisotropic thermal conduction in magnetized
low-collisionality plasmas fundamentally changes the
\citet{Schwarzschild1958} criterion for convective stability
\citep[hereafter Q08]{Balbus2000,Balbus2001,Quataert2008}.  Instead of
convection being driven solely by an inwardly increasing entropy
gradient, low-collisionality plasmas are convectively unstable for
both {\em outwardly} and {\em inwardly} increasing temperature
gradients \citep[Q08]{Balbus2000}.  These instabilities have been
dubbed the magnetothermal instability (MTI; $dT/dr < 0$) and the
heat-flux driven buoyancy instability (HBI; $dT/dr > 0$).

The key driver for both of these instabilities is the highly
anisotropic heat flux established by even a dynamically weak magnetic
field.  In particular, if the electron cyclotron frequency is large
compared to the electron collision frequency, heat flows almost
entirely along magnetic field lines.  However, in the presence of a
magnetic field, the conductivity tensor {\em always} possesses some
degree of anisotropy, even when collisions are rapid.  Thus, any
plasma in which electron conduction is energetically important may be
susceptible to the MTI and HBI.  This motivates us to generalize
previous analyses of conduction-mediated buoyancy instabilities to
collisional plasmas.  We are particularly interested in the possible
application of this physics to the highly conducting plasmas in
neutron stars (NSs) and white dwarfs (WDs) and so we will allow for
the possibility of degenerate plasmas.

A reasonable measure of the collisionality of a plasma is given by the
ratio of the electron cyclotron frequency to the electron collision
frequency, $\omega_g\tau$, where $\omega_g = e B/\gamma m_e c$ is the
Larmor frequency of electrons with Lorentz factor $\gamma$, $B$ is the
magnetic field strength, and $\tau$ is the time between collisions.
To motivate the parameter regime of interest in this paper, we
estimate $\tau$ as the inverse of the electron-ion collision rate,
$\nu_{ei} = n_i \sigma_{ei} v_e$, where $n_i$ is the number density of
ions, $\sigma_{ei}$ is the Coulomb cross section, and $v_e$ is the
(thermal or Fermi) velocity of the electrons.  For the degenerate plasmas of interest in
WDs and NSs, the kinetic energy equals the Fermi energy $E_{\rm F}$.
Setting the cross section to be $\sigma_{ei} = \pi b^2\ln\Lambda$,
where the impact parameter is $b = Ze^2/E_{\rm F}$, $Z$ is the ion's
charge, and $\ln\Lambda$ is the Coulomb logarithm, which is of order
unity in degenerate plasmas \citep{Clayton1983}, we find
\begin{equation}\label{eq:omegatau_deg_nr}
\omega_g\tau \approx  
0.1 \left(\frac{B}{10^8\,{\rm G}}\right)\frac 1 {1 + (\rho_6/\mu_e)^{2/3}}\left(\frac {1} {\ln\Lambda}\right)\frac 1 Z
\end{equation}
where the mass density equals $10^6 \rho_6$ g cm$^{-3}$ and $\mu_e$ is
the electron mean molecular weight.  The factor of $[1 +
(\rho_6/\mu_e)^{2/3}]^{-1}$ in equation (\ref{eq:omegatau_deg_nr})
accounts for the extension to relativistic degeneracy. Equation
(\ref{eq:omegatau_deg_nr}) shows that degenerate plasmas are highly
collisional unless the magnetic field exceeds $\gtrsim 10^{9}\,{\rm
  G}$ (interiors of WDs; $\rho_6 \sim 1$, $\mu_e = 2$) or $\gtrsim
10^{14} \,{\rm G}$ (interiors of NSs; $\rho_6 \sim 10^{8-9}$, $\mu_e
\simeq 10$).  Hence, magnetically driven buoyancy instabilities, if
they exist in weakly magnetized NSs or WDs, operate in the
collisional limit studied in this paper.

The remainder of this paper is organized as follows.  In
\S\ref{sec:MTI calculation}, we present a linear calculation of
buoyancy instabilities in weakly magnetized plasmas allowing for
arbitrary collisionality.  We present order of magnitude estimates of
the growth rates and conditions for instability in the collisional
limit.  We then discuss the application of these instabilities to NSs
and WDs in \S\ref{sec:application}.  We summarize and discuss our
results in \S\ref{sec:discussion}.

\section{Linear Calculation}\label{sec:MTI calculation}

We consider a magnetized fluid with an arbitrary equation of state and
arbitrary collisionality. The basic equations governing the plasma are
the continuity equation,
\begin{equation}\label{eq:continuity}
\frac{\partial \rho}{\partial t} + \boldnabla\cdot\left(\rho\boldv\right) = 0,
\end{equation}
the momentum equation,
\begin{equation}\label{eq:momentum}
\rho\frac{\partial \boldv}{\partial t} + \rho\left(\boldv\cdot\boldnabla\right)\boldv = \frac 1 {4\pi} \left(\boldnabla\times\boldB\right)\times\boldB - \boldnabla P - \rho \boldg,
\end{equation}
the induction equation,
\begin{equation}\label{eq:induction}
\frac{\partial \boldB}{\partial t} = \boldnabla\times\left(\boldv\times\boldB\right),
\end{equation}
and the energy equation, taken here to be an equation for the entropy
of the fluid,
\begin{equation}\label{eq:heat}
\frac P {\Gamma_3 - 1}\left[\frac{\partial}{\partial t} + \left(\boldv\cdot\boldnabla\right)\right]\ln P\rho^{-\Gamma_1}= -\boldnabla\cdot\boldQ,
\end{equation}
where $\rho$ is the density, $\boldv$ is the fluid velocity, $\boldB$
is the magnetic field, $P$ is the pressure, $\boldg$ is the local
gravitational acceleration, $T$ is the temperature, $\Gamma_1 =
\left(\partial \ln P/\partial\ln\rho\right)_s$ and $\Gamma_3 - 1 =
\left(\partial \ln T/\partial \ln \rho\right)_s$ are adiabatic
indices, and $s$ is the specific entropy.  The heat flux, $\boldQ$, is
due to radiative diffusion and conduction and is given by
\begin{equation}\label{eq:old heat flux}
\boldQ = -\chi_{\rm rad} \boldnabla T - \chiperp \left(\boldnabla - \boldb(\boldb\cdot\boldnabla)\right)T - \chipara \boldb\left(\boldb\cdot\boldnabla\right)T,
\end{equation}
where $\boldb = \boldB/B$ is the unit vector in the direction of
$\boldB$ and $B = |\boldB|$.  It is helpful to rewrite equation
(\ref{eq:old heat flux}) as
\begin{equation}\label{eq:heat flux}
\boldQ = -\left(\chi_{\rm rad} + \chiiso\right) \boldnabla T - \chipara' \boldb\left(\boldb\cdot\boldnabla\right)T, 
\end{equation}
where $\chiiso = \chiperp$ is the "isotropic" part of the electron
conductivity and $\chipara' = \chipara - \chiperp$ is then the enhancement
of the electron conductivity along the magnetic field.  From
\cite{Braginskii1965},
\begin{equation}
\chiperp = {\chi \over 1 + (\omega_g \tau)^2} \ \ {\rm and} \ \ \chipara = \chi,
\end{equation}
where $\chi$ is the nonmagnetic thermal conductivity.  For
$\omega_g\tau \gg 1$, the plasma has low collisionality, $\chiperp
\rightarrow 0$, and $\chipara' = \chipara = \chi$.  This is typically
the case for interstellar and intergalactic plasmas and stellar
coronae.  This anisotropic conduction gives rise to the buoyancy
instabilities described in the Introduction.  For $\omega_g \tau \ll
1$, on the other hand, the plasma is highly collisional.  This is
typical of stellar interiors, WDs, and weakly magnetized NSs.  In this
limit, electron conduction is largely isotropic with only a small
anisotropic correction: $\chiperp \simeq \chi$ and $\chipara' \simeq
(\omega_g \tau)^2 \chi$.

Finally, we assume that there may be composition gradients in the
plasma, but that the timescale for the composition to change is long
compared to the other timescales of interest.\footnote{In \S
  \ref{sec:application}, we focus on the application of the
  collisional MTI to the $\rho \sim 10^{4-6}$ g cm$^{-3}$ atmospheres
  of NSs where composition-changing reactions are negligible
  (excluding episodic fusion due to accretion).}  Thus there is no
Lagrangian change in the composition:
\begin{equation}\label{eq:electron}
\frac{\partial\mu_e}{\partial t} + \boldv\cdot\boldnabla\mu_e = 0.
\end{equation}
For concreteness, we have focused on background gradients in the
electron mean molecular weight $\mu_e$, but our calculation can easily
be generalized to allow for other composition gradients (e.g., in the
ion abundance).

We now perform a standard WKBJ analysis on equations
(\ref{eq:continuity})-(\ref{eq:electron}).  We linearize equations
(\ref{eq:continuity}) - (\ref{eq:electron}) using the Boussinesq
approximation as in \citet{Balbus2000} and Q08.  We assume
perturbations of the form $\exp\left(\sigma t -
  i\boldk\cdot\boldx\right)$ and that gravity is in the vertical
direction.  Under these assumptions the perturbed continuity equation
is given by
\begin{equation}\label{eq:incompressibility}
\boldk\cdot\delta\boldv = 0,
\end{equation}
which is the standard incompressibility condition; the perturbed
momentum equation is
\begin{equation}\label{eq:perturbed momentum}
  \sigma\delta\boldv = -i\boldk \frac{\boldB\cdot\delta\boldB}{4\pi\rho} + i \delta\boldB\frac{\boldB\cdot\boldk}{4\pi\rho} - i\boldk\frac{\delta P}{\rho} + \frac{\delta\rho}{\rho} \frac {\boldnabla P} {\rho},
\end{equation}
the perturbed induction equation is
\begin{equation}\label{eq:perturbed induction}
\delta\boldB = i\delta\boldv\frac{\boldB\cdot\boldk}{\sigma},
\end{equation}
and the perturbed heat equation is
\begin{equation}\label{eq:perturbed heat}
\frac P {\Gamma_3 - 1} \left(-\Gamma_1\sigma \frac {\delta \rho}{\rho} + \delta\boldv \cdot \boldnabla \ln P\rho^{-\Gamma_1}\right) = -i \boldk\cdot\delta\boldQ 
\end{equation}
where 
\begin{eqnarray}
\delta\boldQ = - i\chiradiso\boldk\delta T
-\delta\chiiso\boldnabla T - \delta\chipara'\boldb\left(\boldb\cdot\boldnabla T\right)
\nonumber\\ - \chipara'\left[\delta\boldb\left(\boldb\cdot\boldnabla T\right)
 + \boldb\left(\delta\boldb\cdot\boldnabla T\right)\right] - i\chipara'\boldb\left(\boldk\cdot\boldb\delta T\right),
\label{eq:deltaQ}
\end{eqnarray}
and $\chiradiso = \chirad + \chiiso$ is the sum of the perpendicular
electron and radiative conductivities.  In the perturbed heat
equation, we have taken $\delta P \approx 0$ as our implementation of
the Boussinesq approximation.\footnote{Taking the dot product of ${\bf
    k}$ with equation (\ref{eq:perturbed momentum}), it is easy to
  show that $\delta (P + B^2/8\pi)/P \ll \delta \rho/\rho$ for short
  wavelength modes; this suggests that we should take $\delta (P +
  B^2/8 \pi) \approx 0$ in the energy equation as our implementation
  of the Boussinesq approximation.  However, this introduces spurious
  growth and damping terms into the dispersion relation, as we
  confirmed by taking the incompressible limit, i.e., the high $\beta
  \equiv P/(B^2/8\pi)$ limit, of the full non-Boussinesq perturbation
  equations. Thus we implement the Boussinesq approximation by taking
  $\delta P/P \ll \delta\rho/\rho$ in the energy equation.} Note that equation
(\ref{eq:deltaQ}) includes the perturbed conductivities
$\delta\chiiso$, and $\delta\chipara'$.  We ignore perturbations in
the conductivities proportional to $\delta T$ since these are higher
order in $1/kH$, where $H$ is the scale height.  However, this leaves
perturbations with respect to ${\bf B}$
\begin{equation}\label{eq:deltachi}
\delta\chiiso = -\chipara' \left[1 + (\omega_g\tau)^2\right]^{-1} \frac {2\boldb\cdot\delta\boldB}{B},
\end{equation}
and $\delta\chipara' = -\delta\chiiso$.  Using equation
(\ref{eq:deltachi}) in the perturbed heat flux, we find
\begin{eqnarray}\label{eq:perturbed heat flux}
-i\boldk\cdot\delta\boldQ = -\left[\chiradiso k^2 + \chipara'\left(\boldb\cdot\boldk\right)^2\right]\delta T + \nonumber\\
i \chipara'\left[\left(\boldk\cdot\delta\boldb\right)\left(\boldb\cdot\boldnabla T\right) +
\left(\boldk\cdot\boldb\right)\left(\delta\boldb\cdot\boldnabla T\right)\right] -\nonumber\\ i2\frac{\chipara'}{1 + (\omega_g\tau)^2}\frac{\delta\boldB\cdot\boldb} B\left[\boldk\cdot\boldnabla T - (\boldk\cdot\boldb)(\boldb\cdot\boldnabla T)\right].
\end{eqnarray}
We calculate $\delta T$ for any given equation of state,
$P(\rho,T,\mu_e)$, using the Boussinesq approximation ($\delta P = 0$)
via
\begin{equation}\label{eq:p perturbation}
 \delta P = 0 = \left(\frac {\partial P}{\partial \rho}\right)_{T,\mu_e} \delta\rho + \left(\frac {\partial P}{\partial T}\right)_{\rho,\mu_e} \delta T + \left(\frac {\partial P}{\partial \mu_e}\right)_{\rho,T} \delta \mu_e .
\end{equation}
This implies
\begin{equation}\label{eq:deltaT}
\frac{\delta T}{T} = \left[\frac{\partial\ln T}{\partial \ln \rho}\right]_{P,\mu_e}\frac {\delta\rho}{\rho} + \left[\frac{\partial\ln T}{\partial \ln \mu_e}\right]_{P,\rho}\frac {\delta\mu_e}{\mu_e}, 
\end{equation}
where
\begin{eqnarray}\label{eq:dlnTdlnrho}
\left[\frac{\partial\ln T}{\partial \ln \rho}\right]_{P,\mu_e} = -\frac {\rho}{T} \frac{(\partial P/\partial \rho)_{T,\mu_e}}{(\partial P/\partial T)_{\rho,\mu_e}},\\ 
\left[\frac{\partial\ln T}{\partial \ln \mu_e}\right]_{P,\rho} = -\frac {\mu_e}{T} \frac{(\partial P/\partial \mu_e)_{T,\rho}}{(\partial P/\partial T)_{\rho,\mu_e}}.
\end{eqnarray}
The $\delta T$ perturbation has two components: one due to the density
perturbation and one due to the $\mu_e$ perturbation. To calculate
$\delta\mu_e$, we use the perturbed form of equation
(\ref{eq:electron}):
\begin{equation}\label{eq:perturbed electron}
\sigma\delta\mu_e = \delta\boldv\cdot\boldnabla\mu_e.
\end{equation}
The temperature perturbation in equation (\ref{eq:deltaT}) induced by
the density perturbation depends on the thermodynamics of the plasma.
For instance, equation (\ref{eq:dlnTdlnrho}) gives $\left[{\partial\ln
    T}/{\partial \ln \rho}\right]_{P,\mu_e} = -1$ for an ideal gas and
$\delta\mu_e = 0$.  On the other hand, one can show that for a plasma
of non-degenerate ions and degenerate electrons, in which the
electrons dominate the pressure (e.g., in a WD or in the outer parts
of a NS), $\left[{\partial\ln T}/{\partial \ln \rho}\right]_{P} =
-P_e/n_i k_B T$, where $n_i$ is the number density of ions.  The
factor of $Z E_{\rm F}/k_B T$ difference between the ideal gas and the
degenerate plasma arises from the fact that the entropy of the
degenerate plasma is contained in the ions, which do not contribute
significantly to the pressure \citep{Shapiro1986}.  Finally, we note
that at very high densities, when the ``ions'' (protons and neutrons)
themselves are degenerate (e.g., in the core of a NS), the heat
capacity is reduced by an additional factor of $k_B T/E_{\rm F, ion}$,
where $E_{\rm F, ion}$ is the ion Fermi energy.

We now combine the perturbed induction equation (eq.
[\ref{eq:perturbed
  induction}]), the perturbed momentum equation (eq. [\ref{eq:perturbed
  momentum}]), and the incompressibility condition
(eq. [\ref{eq:incompressibility}]) to find:
\begin{equation}\label{eq:delta v}
\left[\sigma^2 + (\boldk\cdot\boldv_{\rm A})^2\right]\delta\boldv = -\sigma\left(\boldg - \frac {\boldk\cdot\boldg}{k^2}\boldk\right)\frac {\delta\rho}{\rho}.
\end{equation}   Combining  equations (\ref{eq:perturbed heat}), (\ref{eq:perturbed heat
  flux}), (\ref{eq:deltaT}), (\ref{eq:perturbed electron}), and (\ref{eq:delta v}) we arrive at the final form of the dispersion relation:
\begin{eqnarray}
-\sigma\sigmabar - \sigma N^2 \frac {k_{\perp}^2}{k^2} + \left(\omega_{\rm iso} + \omega_{\rm cond,\parallel}\right)\nonumber\\
\times\left(\sigmabar \left[\frac {\partial\ln T}{\partial \ln\rho}\right]_{P,\mu_e} + \left[\frac {\partial\ln T}{\partial \ln\mu_e}\right]_{P,\rho} g \frac{\partial \ln\mu_e}{\partial z}\frac{k_{\perp}^2}{k^2}\right)\frac {P_{\rm Th}}{P}\nonumber \\
- \omega_{\rm cond,\parallel} g \frac{\partial\ln T}{\partial z}\left[\frac {k_{\perp}^2}{k^2}\left(1 - 2\Theta b_z^2\right) + 2\Theta\frac{b_x k_x b_z k_z}{k^2}\right] \frac {P_{\rm Th}}{P} \nonumber\\
-2\frac{\omega_{\rm cond,\parallel}}{1 + (\omega_g\tau)^2} g \frac{\partial\ln T}{\partial z} 
\left[\frac{k_z^2}{k^2} \frac {b_x k_x}{\boldk\cdot\boldb} - \frac{k_{\perp}^2}{k^2}\frac{k_z b_z}{\boldk\cdot\boldb}\right]\frac {P_{\rm Th}}{P} = 0
\label{eq:dispersion}
\end{eqnarray}
where $\boldk_{\perp}$ is the component of the k-vector {\it
  perpendicular to $\boldg$}., i.e. $k_{\perp}^2 = k_x^2 + k_y^2$,
\begin{displaymath}
N^2 = g\left(\frac 1 {\Gamma_1}\frac{\partial\ln P}{\partial z} - \frac{\partial\ln\rho}{\partial z}\right) 
\end{displaymath}
is the Brunt-V\"ais\"al\"a frequency, $\Theta \equiv 1 -
(1+[\omega_g\tau]^2)^{-1}$, and $\sigmabar = \sigma^2 + (\boldk\cdot\boldv_{\rm A})^2$.  We have also defined the characteristic  thermal frequencies as
\begin{eqnarray}
\omega_{\rm iso} &=& \frac {\Gamma_1}{\Gamma_3 - 1}  \chiradiso \frac {T} {P_{\rm Th}} k^2, \nonumber\\
\omega_{\rm cond,\parallel} &=&  \frac {\Gamma_1}{\Gamma_3 - 1}  \chipara' \frac {T} {P_{\rm Th}} \left(\boldk\cdot\boldb\right)^2, \nonumber
\end{eqnarray}
where $P_{\rm Th}$ is the thermal component of the total pressure. 

The dispersion relation in equation (\ref{eq:dispersion}) generalizes
the analyses of \cite{Balbus2000,Balbus2001} and Q08 to the case of
arbitrary collisionality and degeneracy. We recover Q08's dispersion
relation by taking the collisionless limit, i.e., $\omega_g \tau \gg
1$, so that $\Theta \rightarrow 1$, and by specializing to an ideal
gas equation of state and no composition gradients.

Motivated by the application to NSs and WDs, we now assume the
following hierarchy of timescales: $t_{\rm dyn} \ll
(\boldk\cdot\boldv_{\rm A})^{-1} \ll \omega_{\rm
  cond,\parallel}^{-1}$, and consider the low frequency (small
$\sigma$) expansion of the dispersion relation in equation
(\ref{eq:dispersion}).  The unstable root of the dispersion relation
is then given by
\begin{equation}\label{eq:lowsigmadispersion}
\sigma \approx \frac{\left(\omega_{\rm iso} + \omega_{\rm cond,\parallel}\right)\left[\Omega_{\rm st}^2 - \Omega_{\rm des}^2\right]}{N^2 {k_{\perp}^2}/{k^2} + (\boldk\cdot\boldv_A)^2},
\end{equation}
where we have defined the stabilizing term as 
\begin{equation}
\Omega_{\rm st}^2 = \frac {P_{\rm Th}}{P}\left\{\left(\boldk\cdot\boldv_A\right)^2
 \left[\frac{\partial\ln T}{\partial \ln\rho}\right]_{P,\mu_e} +  g\frac{\partial\ln\mu_e}{\partial z}\frac{k_{\perp}^2}{k^2}\left[\frac{\partial\ln T}{\partial \ln\mu_e}\right]_{P,\rho}\right\},
\label{eq:stabilizing}
\end{equation}
and the (possibly) destabilizing term as
\begin{eqnarray}
\Omega_{\rm des}^2 \equiv \frac{\omega_{\rm cond,\parallel}}{\omega_{\rm iso} +
  \omega_{\rm cond,\parallel}} g \frac {\partial\ln T}{\partial z} 
\frac {P_{\rm Th}}{P}
\times \nonumber \\
\hspace{2cm} \left[\frac {k_{\perp}^2}{k^2}\left(1 - 2 \Theta b_z^2\right) + 2\Theta \frac {b_zk_zb_xk_x}{k^2} \right.
\nonumber \\
\left. - \frac {2(\boldb\cdot\boldk)^{-1}} {1 + (\omega_g\tau)^2}\left({b_zk_z}\frac{k_{\perp}^2}{k^2} - b_xk_x\frac{k_z^2}{k^2}\right)\right]
\label{eq:destabilizing}
\end{eqnarray}
For $N^2 > 0$ (i.e., a Schwarzschild stable plasma), equation
(\ref{eq:lowsigmadispersion}) gives the instability criterion:
\begin{equation}\label{eq:total stability}
\Omega_{\rm st}^2 - \Omega_{\rm des}^2 > 0.
\end{equation}
Note that our definition of $\Omega_{\rm st}$ is such that
$\Omega_{\rm st}^2 < 0$ since i) $\left[{\partial\ln T}/{\partial
    \ln\rho}\right]_{P,\mu_e} < 0$ and ii) $\partial\ln\mu_e/\partial
z < 0$ for convective stability (in the sense of the Schwarzschild
criterion, $N^2 > 0$).  Equation (\ref{eq:total stability}) is also
the general condition for instability that one can derive by applying
the Routh-Hurwitz criterion as in \citet{Balbus2000}.  Physically,
equation (\ref{eq:total stability}) is the condition that the
destabilizing thermal effects of anisotropic conduction in the
collisional limit exceed the stabilizing effects of magnetic tension
and a stable $\mu_e$ gradient.  The $\mu_e$ gradient is particularly
stabilizing in a degenerate plasma.  First, it increases the magnitude
of $N^2$, thus suppressing the growth rate
(eq. [\ref{eq:lowsigmadispersion}]).  Even more importantly, the
effect of a stabilizing $\mu_e$ gradient is at least a factor of $Z
E_F/k_B T$ {\it larger} than a comparable destabilizing thermal
gradient.  Thus in a degenerate plasma, instability can only arise
when the $\mu_e$ gradient is essentially zero.  For now, we assume
that this is the case.  We will return to this question in \S
\ref{sec:discussion}.

Focusing on the case of zero $\mu_e$ gradient, we find that the
condition for instability is
\begin{equation}\label{eq:stability}
(\boldk\cdot\boldv_A)^2 - \Omega_{\rm des}^2 \left[\frac{\partial\ln T}{\partial \ln\rho}\right]_{P,\mu_e}^{-1} < 0.
\end{equation}
For a
collisional plasma, i.e., $\omega_g \tau \ll 1$, the instability
criterion becomes, at the order of magnitude level,
\begin{equation}\label{eq:OOMinstability1}
(\boldk\cdot\boldv_{\rm A})^2 \lesssim \omega_g^2 \tau^2 \frac {\chi_{\rm cond}(\boldk\cdot\boldb)^2}{\chiradiso k^2 } \left|g\frac{\partial\ln T}{\partial z}
\left[\frac {\partial\ln T}{\partial \ln \rho}\right]_{P,\mu_e}^{-1}\right|.
\end{equation}
Taking $(\boldk\cdot\boldb)^2 \sim k^2$, we can rewrite equation (\ref{eq:OOMinstability1}) as
\begin{equation}\label{eq:OOMinstability}
 (kH)^2 \lesssim \frac {4\pi\rho e^2}{m_e^2 c^2}\tau^2 \frac {\chi_{\rm cond}}{\chiradiso} gH^2 \left|\frac{\partial\ln T}{\partial z}
\left[\frac {\partial\ln T}{\partial \ln \rho}\right]_{P,\mu_e}^{-1}\right|,
\end{equation}
where $H$ is a typical vertical length-scale in the system, i.e., the
pressure scale height.  Note that the instability criterion in
equation (\ref{eq:OOMinstability}) is independent of B because the
$B^2$ that arises from $v_{\rm A}^2$ is canceled by the factor of
$B^2$ from $\omega_g^2$.  The order of magnitude growth time for the
fastest growing mode can now be determined from equation
(\ref{eq:lowsigmadispersion}):
\begin{equation}\label{eq:OOMgrowth}
t_{\rm gr} \sim \omega_{\rm cond,\parallel}^{-1}  \sim t_{\rm Th} \left[(\omega_g\tau)(k H)\right]^{-2},
\end{equation}
where the largest allowed value of $k H$ is given by equation
(\ref{eq:OOMinstability}) and $t_{\rm Th} \sim H^2/\chiiso$ is the
local thermal time. Unlike the instability criterion, the growth time
does depend on $B$, with $t_{\rm gr} \propto B^{-2}$ for a collisional
plasma.

Before applying our results to NSs and WDs, we comment on some of the
physics of the pure MTI ($b_x = 1$) and HBI ($b_z = 1$) instabilities
in the collisional limit.  The primary difference in our dispersion
relation relative to previous works is the fact that the conductivity
depends on B, which introduces terms $\propto \delta\chi$.  For the
pure HBI case, however, where $b_z = 1$ and $\partial T/\partial z >
0$, we find that all terms associated with $\delta\chi$ exactly
cancel.  Thus the basic physics of the HBI in Q08 is recovered in the
collisional limit, albeit with modifications due to isotropic heat
transport associated with $\omega_{\rm iso}$.  On the other hand, the
pure MTI case, in which $b_x = 1$, has somewhat different physics in
the collisional limit.  For a low collisionality plasma, the growth
rate in the pure MTI limit, for our ordering of timescales, is given
by
\begin{equation}
\sigma \approx \omega_{\rm cond,\parallel} \frac{g |\partial\ln T/\partial \ln z|}{N^2}.
\end{equation}
By contrast, the growth rate in a collisional plasma is
\begin{equation}\label{eq:ideal collisional MTI}
\sigma \approx \omega_{\rm cond,\parallel} \frac{g|\partial\ln T/\partial \ln z|}{N^2}\left( 1 + \frac {2k_z^2}{k_\perp^2}\right).
\end{equation}
Equation (\ref{eq:ideal collisional MTI}) shows that the driving of
the MTI is {\it enhanced} in a collisional plasma relative to what a
simple extrapolation of the low-collisionality result might suggest.
Namely, we find that including the fact that the magnitude of the
anisotropy of the heat flux on B in our calculation increases the
growth rate relative to a calculation that considers only a fixed degree
of anisotropy (independent of B).  Note, however, that, despite the
appearance of equation (\ref{eq:ideal collisional MTI}), we require
$\boldk\cdot\boldb \neq 0$ for instability because $\omega_{\rm
  cond,\parallel} \propto (\boldk\cdot\boldb)^2$; for the pure MTI
limit, $b_x = 1$ and thus growth still requires $k_\perp \ne 0$ in a
collisional plasma.  Thus the additional driving does not change the
conditions under which there is growth (e.g., $dT/dz < 0$, weak field,
and $k_\perp \ne 0$), only the growth rate.

The enhanced driving of the MTI in a collisional plasma arises because
of the dependence of the conductivity on $B$; this is analogous to a
$\kappa$ mechanism in stellar oscillation theory \citep{Cox1983}.  To illustrate this
point, we study the perturbed heat fluxes for the pure MTI case, which
are given by
\begin{eqnarray}
\delta\boldQ_x &=& -i\chiradiso k_x \delta T - i\chipara' k_x \delta T -
k_x \chipara' \xi_z \frac {\partial T}{\partial z},\label{eq:Qx} \\
\delta\boldQ_z &=& -i\chiradiso k_z \delta T 
+ 2 \chipara' k_z \xi_z \frac {\partial T}{\partial z}, \label{eq:Qz}
\end{eqnarray}
where $\boldzeta = \delta \boldv/\sigma$.  In the low-collisionality
case studied by Q08, the $\delta\boldQ_x$ term is the same (except for
the inclusion of a perturbed isotropic heat flux), while
$\delta\boldQ_z = 0$.  The first term on the RHS of equation
(\ref{eq:Qz}) is the perturbed isotropic diffusion of heat due to
the perturbed temperature.  The second term is the key new physics: it
represents the change in the vertical heat flux due to the perturbed
{isotropic} conductivity that arises from the perturbed magnetic
field, i.e., $\delta\chiiso \propto \delta B$.  The physical
interpretation is that as a fluid element in an MTI unstable situation
is perturbed, the strength of the magnetic field increases.  This
makes conduction more anisotropic, which increases the conductive
heating/cooling of the fluid.  This in turn leads to a buoyant
response followed by a greater bending of field lines, greater
anisotropic conductivity, and yet more heating.  This additional
driving due to the dependence of the conductivity on B is the origin
of the modest enhancement in the growth of the MTI in a collisional
plasma.

\subsection{The Low-collisionality Limit}
\label{sec:collisionless}

For sufficiently strong magnetic fields, $\omega_g \tau \gtrsim 1$,
and the low-collisionality limit studied by \cite{Balbus2001} and
\cite{Quataert2008} is appropriate, rather than the collisional limit
highlighted here.  For the applications of interest in this paper,
$t_{\rm dyn} \ll t_{\rm Th}$, in which case the order of magnitude
instability criterion in the low-collisionality limit can be
determined from equations (\ref{eq:stabilizing})-(\ref{eq:total
  stability}) with $\theta \rightarrow 1$.  This gives
\begin{equation}\label{eq:OOMinstabilityB}
(k H)^2 \lesssim \frac {\chi_{\rm cond}(\boldk\cdot\boldb)^2}{\chiradiso k^2 } {g H^2 \over v_A^2} \left|\frac{\partial\ln T}{\partial z}
\left[\frac {\partial\ln T}{\partial \ln \rho}\right]_{P,\mu_e}^{-1}\right|.
\end{equation}
Since equation (\ref{eq:OOMinstabilityB}) implies that $k_{max}
\propto B^{-1}$ in the low-collisionality limit, decreasing the
magnetic field strength increases the number of unstable modes that
can fit in the system.  This trend continues until the magnetic field
is sufficiently weak that $\omega_g \tau \lesssim 1$ and the
collisional limit, rather than the collisionless limit, is
appropriate.  In the collisional limit, the magnetic field dependence
of the thermal conductivity cancels that of the Alfv\'en speed and
$k_{max}$ is independent of $B$ (eq. [\ref{eq:OOMinstability}]). The
collisional limit thus allows the largest range of wavelengths to be
unstable.

The order of magnitude growth time in the low-collisionality limit is
given by
\begin{equation}\label{eq:OOMgrowthB}
t_{\rm gr} \sim t_{\rm Th} \left(k H\right)^{-2}.
\end{equation}
The bound on k from equation (\ref{eq:OOMinstabilityB}) implies that
the fastest growing mode in the low-collisionality limit has a growth
time $t_{\rm gr} \propto B^2$.  By contrast, in the collisional limit,
the fastest growing mode has $t_{\rm gr} \propto B^{-2}$
(eq. [\ref{eq:OOMgrowth}]).  Thus the growth time of the MTI and/or
HBI is minimized when $\omega_g \tau \sim 1$, i.e., at the transition
between the collisional and low-collisionality limits.

\section{Applications to Compact Objects}\label{sec:application}

The interiors of main sequence stars have $\omega_g \tau \ll 1$ and
are thus a possible site for the application of the collisional MTI
and HBI instabilities.  However, it is straightforward to show that
the order of magnitude instability condition
(eq. [\ref{eq:OOMinstability}]) becomes $k H \lesssim 10^{-3}$ in the
solar interior; thus no unstable modes can fit in the star.  A more
promising site for the application of these instabilities is to WDs
and NSs, since thermal conduction dominates over radiative diffusion
throughout much of the star.  Provided that the instability criterion
(eq. [\ref{eq:OOMinstability}]) can be satisfied and the growth time
(eq. [\ref{eq:OOMgrowth}]) is sufficiently fast, the collisional MTI
and HBI may be able to reorient small scale magnetic fields in NSs and
WDs or provide a mechanism by which the thermal energy can be tapped
to amplify the local magnetic field.

\subsection{The Cooling Cores of WDs and NSs}

WDs and NSs are born hot, with significant temperature gradients left
over from the post-main sequence evolution of their progenitors.
After about a thermal time, the core becomes isothermal, with the
temperature of the core then decreasing in time as energy is lost
through the thermally insulating surface layers.  A natural location
to study the collisional MTI and HBI is in the cooling core of a WD or
NS before it becomes isothermal.\footnote{Neutrino cooling at the
  center of the progenitor star typically leads to a temperature
  maximum at a non-zero radius in newly formed WDs, so that the core
  has both $dT/dr < 0$ and $dT/dr > 0$.  The WD is thus in principle
  susceptible to both the collisional MTI and HBI.  The same is true
  for cooling NSs.} The condition for instability given by equation
(\ref{eq:OOMinstability}) can be utilized to estimate the most
unstable wavelength; we assume $g \approx GM/R^2$ and $d\ln T/dz \sim
R^{-1}$ for simplicity.  Using the electron-ion collision time from
\citet{Yakovlev1980} -- which is consistent with equation
(\ref{eq:omegatau_deg_nr}) -- we find that the unstable modes in the
core of a nonrelativistic degenerate WD must satisfy
\begin{equation}
  kR \lesssim 13 \, Z_6^{-3/2} \ln \Lambda^{-1} M_{0.6}^{2/3} R_9^{-1} T_8^{1/2} \mu_e^{1/3}, \label{kmaxWD}
\end{equation}
where $T = 10^8 T_8$ K is the initial temperature of the core of the
WD, $R = 10^9 R_9$ cm is its radius, and $M = 0.6 M_{0.6} M_\odot$ is
its mass.

In the core of a NS, neutron and proton degeneracy pressure play an
important role.  The heat capacity of the ions is thus suppressed by a
factor of $\sim (k_BT/E_{\rm F,ion})^2$, which reduces the
effectiveness of the MTI and HBI relative to WDs. We find that modes
are unstable in the core of a NS provided that
\begin{equation}
 kR \lesssim 0.2 R_{6}^{3/2} M_{1.4}^{-1/6} \mu_{e,10}^{5/6} T_8 Z^{-3/2} \ln \Lambda^{-1}.\label{kmaxNS}
\end{equation}
This shows that the cores of weakly magnetized NSs are unlikely to be
unstable to the MTI or HBI because the unstable modes have wavelengths
larger than the size of the star.  This estimate did not include the
isotropic conductivity due to neutrons in the core of the NS, which
may be substantial \citep{Baiko2001,Gnedin2001}; this would further
inhibit the instability.  Strongly magnetized NSs with $B \gtrsim
10^{14}$ G (i.e., magnetars) have $\omega_g \tau \gtrsim 1$
(eq. [\ref{eq:omegatau_deg_nr}]) and thus the collisional limit in
equation (\ref{kmaxNS}) does not apply.  However, as shown in \S
\ref{sec:collisionless} the collisional limit allows the largest range
of wavelengths to be unstable.  Thus the MTI and HBI are even less
likely to be important in the cores of newly formed magnetars.

Equation (\ref{kmaxWD}) indicates that a cooling WD is unstable to the
collisional MTI and HBI for long wavelengths $\gtrsim 0.1$ R.
However, the instability in the core of the WD is driven by the {\it
  initial} temperature gradient, which vanishes after one thermal
time, $t_{\rm Th} \sim \omega_{\rm cond}^{-1}$, after which the core
becomes isothermal.  Thus growth is only important if $t_{\rm gr}
\lesssim t_{\rm Th}$.  Using the estimate of $t_{\rm gr}$ in equation
(\ref{eq:OOMgrowth}), we find the minimum magnetic field required for
growth:
\begin{equation}\label{BminWD}
 B \gtrsim 5\times 10^8 Z_6^{5/2} \ln \Lambda^2 M_{0.6}^{-2/3} R_9 T_8^{-1/2} \mu_{e,2}^{-1/3}\,{\rm G}. 
\end{equation}
The strongest magnetic fields measured in WDs are $\sim 10^9$ G
\citep{Schmidt2003,Vanlandingham2005}. The large magnetic field
required for significant growth in the core of a WD
(eq. [\ref{BminWD}]) implies that the MTI and HBI are only potentially
important for modifying the magnetic fields of the most strongly
magnetized WDs.

\subsection{The Collisional MTI in the Oceans of Neutron Stars}

Even after the core of a WD/NS has become isothermal, a temperature
gradient can persist in the atmosphere of the star.  The MTI can
operate in this outer envelope.  We focus on the application to NSs
because our estimates indicate that the growth times exceed the Hubble
time in the atmospheres of WDs.

To study the stability properties of NS envelopes we first construct
hydrostatic constant-flux equilibrium models of a NS in plane parallel
geometry.  We focus on weakly magnetized, highly ionized plasmas and
assume that the radiative opacity is given by a combination of
free-free absorption and Thomson scattering, while the conductive
opacity is given by electron-ion scattering
\citep{Ventura2001,Brown2002,Chang2003}, following
\cite{Yakovlev1980}.\footnote{We have not included magnetic
  corrections to the opacity in the atmosphere model.}  We have also
tried more modern conductivities such as those of \cite{Potekhin1999};
the conductivities are consistent to about 30\% and the growth times
to a factor of $\sim 2$, which is sufficient for our purposes.

\begin{figure*}
\begin{center}
\includegraphics[width=.4\textwidth]{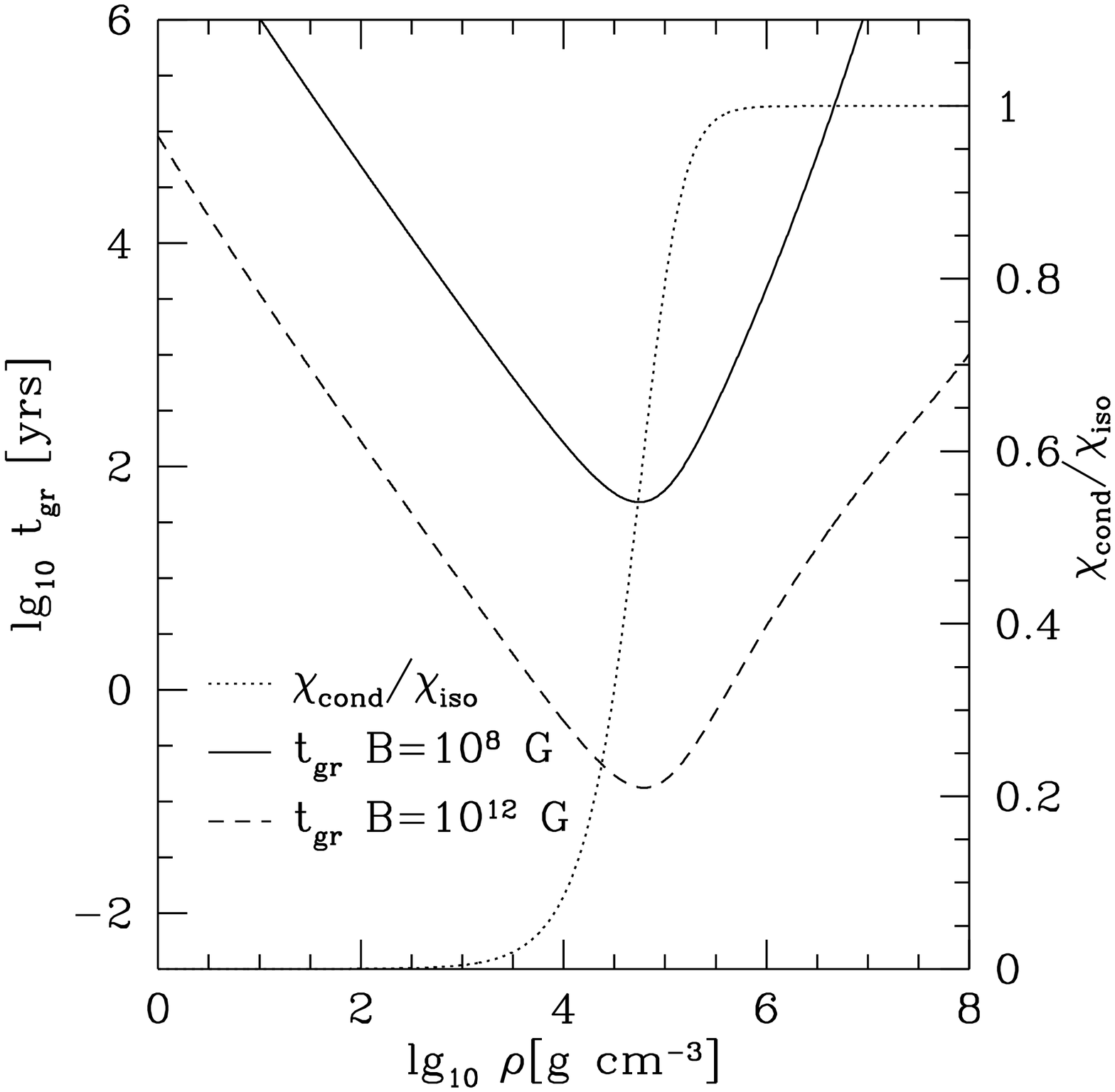}
\includegraphics[width=.4\textwidth]{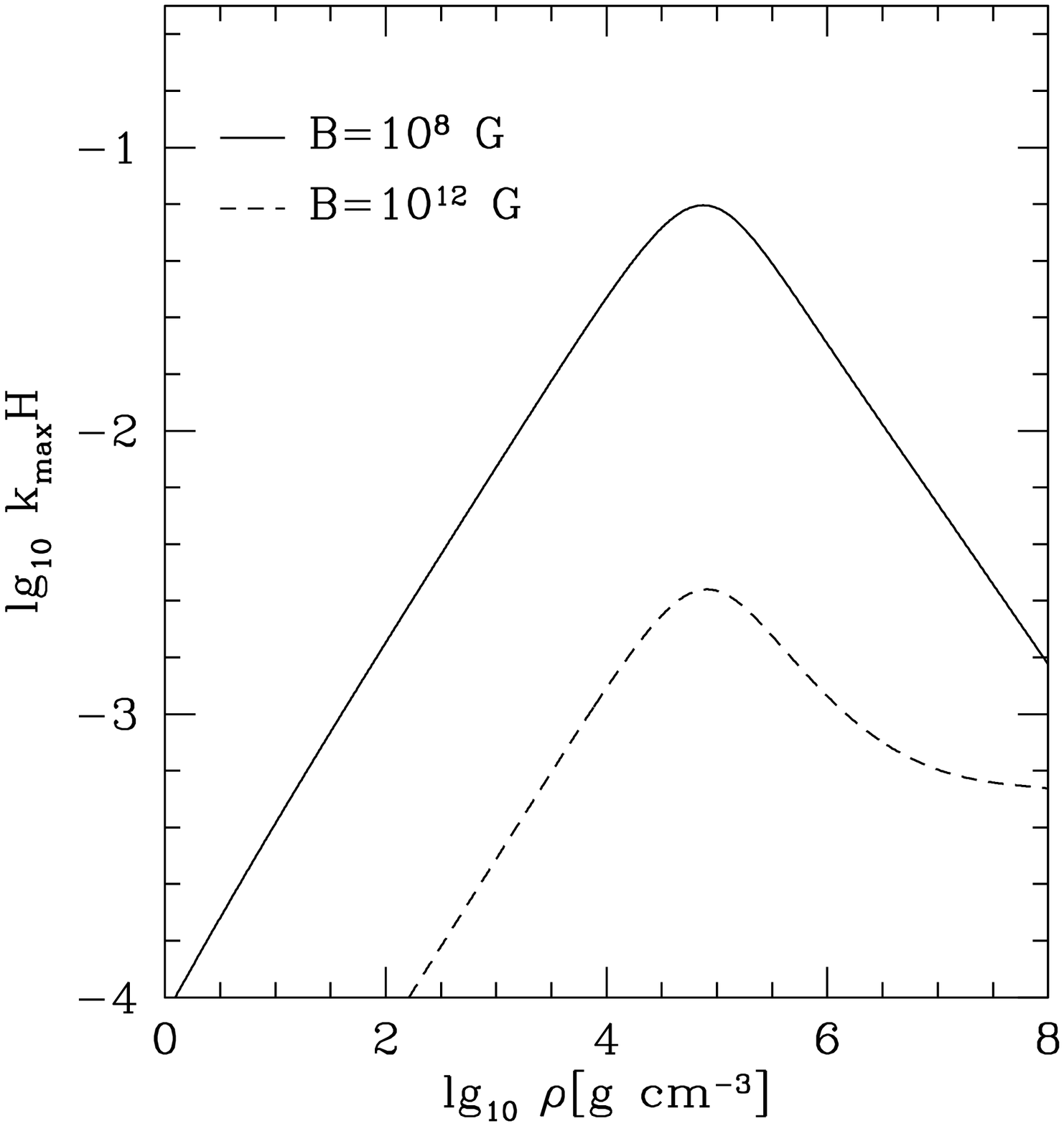}
\end{center}
\caption{{\it Left:} Order of magnitude growth time (solid line) for
  the collisional MTI as a function of density in the envelope of a NS
  with $T_{\rm eff} = 10^6\,{\rm K}$ and $B=10^8$ G.  The minimum
  growth time (maximum growth rate) is $\approx 6\,{\rm yrs}$ and
  occurs where the opacity transitions from radiative to conductive
  (dotted line).  In the collisional limit, $t_{\rm gr} \propto
  B^{-2}$ (eq. [\ref{eq:OOMgrowth}]). {\it Right:} The maximum
  wavevector (minimum wavelength; eq. [\ref{eq:OOMinstability}])
  relative to the pressure scale height $H$ as a function of $\rho$
  for the same model. In the collisional limit, $k_{\rm max}$ is
  independent of B.} \label{fig:NS} \end{figure*}

In the left panel of Figure \ref{fig:NS} we plot the order of
magnitude growth time for the fastest growing mode, $t_{\rm gr}$ (eq.
[\ref{eq:OOMgrowth}] and [\ref{eq:OOMgrowthB}]), as a function of
density in the atmosphere of a 1.4 $\Msun$, 10 km NS with $T_{\rm eff}
= 10^6\,{\rm K}$ for $B=10^8$ G (solid line) and $B=10^{12}$ G (dashed
line). Recall that the growth time in the collisional limit scales as
$t_{\rm gr} \propto B^{-2}$, while in the collisionless limit it
scales as $t_{\rm gr} \propto B^{2}$. The minimum growth time $t_{\rm
  gr,min}$ is $\sim 50$ yrs ($\sim 0.1$ yrs) for $B=10^8$ G ($10^{12}$
G) and occurs at the ``sensitivity strip'' \citep{Ventura2001} where
the radiative envelope transitions to the isothermal interior, i.e.,
where $\chi_{\rm cond}/\chi_{\rm iso} \approx 0.5$.  This is not
surprising since at smaller radii (higher $\rho$), there is no
significant temperature gradient to drive the instability, while at
larger radii (lower $\rho$), the opacity is largely radiative. The
Coulomb logarithm $\Gamma$ is $< 173$ where the growth peaks, so that
the instability occurs largely in the liquid atmosphere, not the
crust.

In the right panel of Figure \ref{fig:NS}, we plot the most unstable
wavevector (eq. [\ref{eq:OOMinstability}] and
[\ref{eq:OOMinstabilityB}]) for $B=10^8$ G and $10^{12}$ G,
respectively. The most unstable wavelength is larger than the scale
height by at least a factor of 10 for the collisional case ($B = 10^8$
G) and by a factor of $\gtrsim 300$ for the low collisionality case
($B = 10^{12}$ G).  Unlike in the core of the star, this does not
completely preclude the MTI from acting because the horizontal
wavelength of the mode can be much larger than the vertical wavelength
in a thin atmosphere.  However, because the tension force is $\propto
{\bf k} \cdot {\bf B}$, Figure \ref{fig:NS} does indicate that the
instability is most likely to act when the background magnetic field
is relatively perpendicular to gravity, so that nearly horizontal
perturbations do not significantly bend the magnetic field lines.
Moreover, in \S \ref{sec:collisionless} we showed that the most
unstable wavelength for the collisional MTI, which is independent of
B, is smaller than the most unstable wavelength for the MTI in the
low-collisionality limit.  Thus the $B = 10^8$ G result for $k_{\rm
  max} H$ in Figure \ref{fig:NS} is the shortest wavelength for the
MTI in any NS atmosphere.  Since $k_{\rm max} H \lesssim 0.1$ even in
this case, magnetic tension severely limits the extent to which the
MTI can operate in the atmospheres of NSs.  In particular, even if the
MTI is present, tension will significantly modify its nonlinear
evolution relative to what one would find for $k H \gg 1$: it will be
difficult for the field to become relatively radial, as occurs in
local simulations of the MTI in the low-collisionality limit
\citep{Parrish2007}.

In spite of the suppression of the MTI by magnetic tension, it is
nonetheless useful to study a range of NS atmospheres, parameterized
by their effective temperatures $T_{\rm eff}$, to see how the
properties of the instability change in different NSs.  We show these
results in Table 1 for $B = 10^8$ G, i.e., in the collisional limit.
We find that the characteristic instability time is always short
($\lesssim 10^2$ yrs) and that the region of maximum growth is always
a liquid ($\Gamma < 173$).  For comparison to our estimated growth
times, the typical thermal time of a NS at $\sim 10^6$ K is $\approx
10^6$ yrs.\footnote{This is primarily set by neutrino cooling. Once
  neutrino cooling ceases, the thermal time is $\gtrsim 10^7$ yrs.}
Thus, at a given phase in the evolution of a cooling NS, it is
unstable to the MTI on a timescale that is very short compared to the
thermal time on which the properties of the NS evolve.  The biggest
caveat is that, as discussed above, the most unstable wavelength at
the location where the growth peaks (eq. [\ref{eq:OOMinstability}]) is
$k_{\rm max} H \sim 0.01-0.1$ (Table 1); $k_{\rm max} H$ is the
largest, and thus growth is the most likely, for younger, hotter NSs.

\begin{table}
 \centering\label{table}
 \begin{minipage}{60mm}
   \caption{Table of MTI properties in NS atmospheres with different
     effective temperatures $T_{\rm eff}$.  The columns include the
     effective temperature, minimum growth time for the MTI
     (eq. [\ref{eq:OOMgrowth}]), the density at which the fastest
     growth occurs, the most unstable wavelength in units of the local
     pressure scale height (eq. [\ref{eq:OOMinstability}]), and the local Coulomb parameter. Units are   K for $T_{\rm eff}$, yrs for $t_{\rm gr, min}$ and g cm$^{-3}$   for $\rho_{\rm min}$.  All models are for a 1.4 $M_\odot$, 10 km
     NS with B = $10^8$ G.  $t_{\rm gr} \propto B^{-2}$ in the
     collisional limit.}
  \begin{tabular}{rrrrr}
  \hline
   $T_{\rm eff}$ &  $t_{\rm gr,min}$  & $\rho_{\rm min}$ & $k_{\rm max} H$ & $\Gamma_{\rm min}$ \\
\hline
   $\times 10^5$ &                    & $\times 10^4$    &                         &                   \\
\hline
        2        & 549               & 0.089            & 0.00653                  & 64.3            \\
        4        & 180               & 0.540            & 0.0172                   & 44.0           \\
        6        & 96.9               & 1.53             & 0.0302                   & 35.1          \\
        8        & 64.2               & 3.15             & 0.0449                   & 29.9          \\
        10       & 47.9               & 5.53             & 0.0609                   & 26.3           \\
        12       & 38.5               & 8.73             & 0.0779                   & 23.7             \\
        14       & 32.8               & 12.8             & 0.0957                   & 21.7         \\
        16       & 29.0               & 17.3             & 0.113                    & 20.0          \\
        18       & 26.6               & 22.8             & 0.132                    & 18.6           \\
	20       & 24.9               & 29.1             & 0.151                    & 17.5          \\
\hline
\end{tabular}
\end{minipage}
\end{table}

\section{Discussion}\label{sec:discussion}

We have studied buoyancy instabilities due to anisotropic heat
conduction in collisional, degenerate plasmas, i.e., when the electron
collision frequency is large compared to the electron cyclotron
frequency.  Although heat conduction is nearly isotropic in this
limit, i.e., it is nearly independent of B, the small residual
anisotropy drives convective instabilities analogous to the MTI
\citep{Balbus2000} and HBI (Q08) that have been studied previously in
low-collisionality plasmas. The physics of the two instabilities is
essentially the same in a collisional plasma, although there is
additional driving of the MTI due to the magnetic-field dependence of
the thermal conductivity (see the discussion near eq. [\ref{eq:ideal
  collisional MTI}]).

In a low-collisionality plasma, the condition for the MTI and/or HBI
to grow depends on the magnetic field strength $B$, with magnetic
tension suppressing the instability for larger $B$ (\S
\ref{sec:collisionless}).  The growth time of the instability also
depends on $B$, with the fastest growing mode having $t_{\rm gr}
\propto B^2$.  By contrast, in a collisional plasma, the competition
between tension and the destabilizing effects of a magnetic-field
dependent thermal conductivity leads to an instability criterion that
is independent of $B$ (eq. [\ref{eq:OOMinstability}]); the growth time
in the collisional limit depends on $B$, however, with $t_{\rm gr}
\propto B^{-2}$.  Magnetic tension thus has the smallest effect on the
MTI and HBI for marginally collisional plasmas in which the electron
collision frequency is comparable to the electron cyclotron frequency.

The collisional MTI and HBI can in principle operate in the cores of
young WDs and NSs and in the outer envelopes of NSs (\S
\ref{sec:application}).  However, magnetic tension and the low
specific heat of a degenerate plasma severely limit the importance of
the MTI and HBI in these environments.  The MTI and HBI can undergo
several e-foldings in the core of a young, high magnetic field WD
($\gtrsim 5 \times 10^8$ G), before thermal conduction wipes out the
temperature gradient that drives the instability in the first place
(eq. [\ref{BminWD}]); however, only a small fraction of WDs have
magnetic fields this strong \citep{Vanlandingham2005}.  We find that
there is very unlikely to be analogous growth of the MTI or HBI in the
cores of young NSs (eq. [\ref{kmaxNS}]).  However, the MTI may be
present in the liquid oceans of NSs (see Fig. \ref{fig:NS}), where a
finite temperature gradient persists even after the core has become
isothermal.  The most promising candidates are young (hot), weakly
magnetized ($B \lesssim 10^9$ G) NSs.  Even in this case, however, the
fastest growing modes have $k_{\rm max} H \sim 0.1$ and are thus
restricted to very long wavelengths (Table 1).  For more strongly
magnetized NSs, in which conduction is more anisotropic, growth in the
outer atmosphere is {\it less likely} because of the increased
stabilization by magnetic tension.

Given these results, one interesting possibility is that the
collisional MTI could contribute to amplifying and/or modifying the
magnetic fields in the atmospheres of weakly magnetized NSs such as
millisecond pulsars or NSs in low-mass x-ray binaries.  In these
systems, it is believed that an initially strong birth magnetic field
either decayed away or was buried by accretion, leaving the NS with a
weak field that is potentially unstable to the MTI.

However, it is unclear whether the collisional MTI will grow
significantly in this context and, if it does, how it will saturate.
In weak-field simulations of the MTI in low-collisionality plasmas,
the magnetic field is amplified by a factor of $\sim 10-30$ and the
instability saturates by re-orienting the magnetic field to be largely
in the direction of the background gravitational field
\citep{Parrish2005,Parrish2007}.  In the oceans of NSs, however, the
instability is restricted to very long wavelengths, larger than the
local pressure scale height of the atmosphere (Fig. \ref{fig:NS} and
Table 1).  Thus the instability is most likely to act in regions where
the local magnetic field is relatively perpendicular to gravity, so
that modes with long horizontal wavelengths do not significantly bend
the magnetic field.  Given these stringent requirements for growth, it
is difficult to see how the MTI can significantly change the structure
of the NSs magnetic field.  Nonlinear simulations are required to
assess this in detail and to determine if the additional driving due
to the magnetic field-dependent thermal conductivity modifies the
nonlinear evolution of the MTI in collisional plasmas.

A background gradient in the composition, in particular in the
electron mean molecular weight $\mu_e$, can also act to stabilize the
MTI and does so particularly effectively in a degenerate plasma (see
the discussion below eq. [\ref{eq:total stability}]).  In the part of
a NSs atmosphere where the instability is the most powerful, however,
we do not expect a significant $\mu_e$ gradient.  The diffusion time
through this region is incredibly fast, $\sim 0.1-10$ s at $\rho \sim
10^5\,{\rm g\,cm}^{-3}$ \citep{Chang2004}.  Thus a significant $\mu_e$
gradient is only likely if the growth happens to occur near a
compositional discontinuity in the NS.  This is not impossible but is
unlikely to generically be the case.

\section*{Acknowledgments}

We thank Tony Piro and Ian Parrish for useful discussions.  We thank 
the referee, Steve Balbus, for a critical reading of this manuscript. P.C. is
supported by the Theoretical Astrophysics Center at UC Berkeley.
E.Q. is supported in part by the David and Lucile Packard Foundation
and NSF-DOE Grant PHY-0812811.

\bibliographystyle{mn2e} 
\bibliography{MTI}

\end{document}